# Cognitive Routing with Stretched Network Awareness through Hidden Markov Model Learning at Router Level

T R Gopalakrishnan Nair
Director, Research and Industry Center, DSI & Department Of Computer Science and Engineering, Dayananda Sagar Institution, Bangalore. India. Email: trgnair@yahoo.com

M Jayalalitha,
Asst. Professor, Research and Industry Center, DSI & Department of Telecommunication Engineering Dayananda Sagar Institution, Bangalore, India. Email: mjaya_lalitha@yahoo.co.in

Abhijith S,
Research and Industry Center Department Of Computer Science and Engineering, Dayananda Sagar Institution, Bangalore. India. Email abhijiths85@gmail.com

## Abstract

**The routing of packets are generally performed based on the destination address and forward link channel available from the instantaneous Router without sufficient cognizance of either the performance of the forward Router or forward channel characteristics. The lack of awareness of forward channel property can lead to packet loss or delayed delivery leading to multiple retransmissions or routing to an underperforming pathway. This paper describes an application of Cognitive Network to improve the network performance by implementing a Hidden Markov Model (HMM) algorithm for learning and predicting the performance of surrounding routers continuously while a routing demand is initiated. The cognition segment/domain of every router can gain knowledge about the quality of forward network. The information of the current network conditions is shared between routers by the 'Forward Channel Performance Index' FCPI. This enables complete cognition of surroundings and efficient delivery of messages in various paradigms of performance.**

## 1. Introduction

A cognitive network is composed of many elements that, through learning and reasoning, dynamically adapt to varying network conditions in order to optimize end to end delivery capability[1][9]. The cognition properties of spatially and temporally forward positioned channels gives the routers discussed here a network-wide scope. This is realized by the sharing of information, which differentiates it from conventional routers that have only a local and single element of scope. With this, the network can become a self-learning and forward looking/prediction system that will discover the anomalies/problem in the network and reconfigure itself against those anomalies without affecting the end to end user goals [9].

Cognition is used in association with a capability to observe channels to make behavioral adjustments. It also receives feedback from other Routers while learning. The elements of this cognitive network are capable of assembling and assimilating the information from surroundings. It helps them to predict the forward behavior of the network based on the past and current states. The performance parameters observed in a network node are collected and uploaded into the network by each cognitive element for decision-making. The decision-making process uses reasoning to determine the next set of actions that can be implemented in the network. Cognitive routers uses the learning mechanism of Hidden Markov Model (HMM) to become proactive rather than reactive to adjust to the delivery problems that may occur and predict the next level network conditions [1]. It enables this Cognitive network to be a forward looking one.

## 2. **Routing Challenges**

The scope of current Routers in cognition is limited. Their response to crisis or other undesirable network conditions are not good enough and they are unable to make intelligent adaptations. They have localized, individual network intentions, unaware of the network conditions/parameters experienced by other Routers in its neighborhood. This inevitably leads to undesirable features such as Congestion, packet loss, retransmission of lost packets, long delay in the packet transmission and sometimes inferior QoS during real time communication services which demand assured bandwidth in communication links and in the Router capabilities itself.

Currently, the transmission of packets is carried out with a collection of router peers where each router has a limited view of the network in real-time, restricted to the output or immediate connection available to an adjacent router. It blindly forwards the message to outgoing links. If a packet does not reach its destination, it is retransmitted on either the same link or other alternative path. This can be one of the main challenges that inhibit carriers from guaranteeing QoS for real-time data services over the Internet.

## 3. Cognition in network

Cognitive Networks are capable of self-learning, predicting the future behavior and reconfiguring itself based on the current dynamic network environment. Learning and predicting operation in a router depends upon the data like the performance Index experienced by future Routers for the messages. For this purpose, all the Router's forward conditions are sent to its neighboring Routers. Information is shared among Routers by sending Cognitive Packets. It takes place at a selected large interval of time which demands negligible channel bandwidth compared to the regular packets carrying information, thus minimizing the demand of capacity for cognition activity in the network. The cognitive router suitable for this type of network will have a cognitive domain and a communication domain inside them.

Cognition domain is not a separate management system but it is an inherent very low frequency management activity that works above the ordinary routing activity. It generally guides the routing operation of the router as shown Figure 1. Any catastrophe on end to end delivery, learned through awareness of external multilayer network hop points will be enabling a rerouting decision in the high speed routing taking place in routing domain. It has to be necessarily an insider to do the multi path learning.

Cognitive layer is however separated from ordinary routing function. The planning, learning, reasoning activities are envisaged to be done in a different processor and they communicate to router administrator. The computational complexity for computing HMM and Forward Channel Performance Index is dealt with at a low frequency compared to routing activity.

## 4. **Elements of a Cognitive Node**

The elements of a cognitive router with stretched environment awareness are shown in Figure.1. Cognitive configuration is a feedback driven structure where constant sensing and analysis of surrounding network is performed.

The information for intelligence building i.e. the cognitive information about the current network condition is collected from inputs and is fed into Planning and Learning tasks in the Router. The output of Learning and planning units are fed in to the reasoning unit named the Advisor and the guidance from the advisor for the routing is handed over to the router administrator at every instant of fresh routing demand.

Figure.1. Elements of Cognitive Node

In this Cognitive Network, the Router is designed to have two domains as shown in the Figure 2.

1) Cognitive Domain
2) Communication Domain.

### 4.1.1 Cognitive Domain

Cognitive Domain has both receiving and sending operations in the form of exchanging the information by Cognitive packets. The Cognitive Domain Learning unit receives network information as a Forward Channel Performance Index (FCPI) in the form of 8 bit values for each neighboring Routers through the Cognitive Packets.

Figure 2. Components of a Cognitive Router

In the cognition domain, the prediction of future conditions of other Routers is done by Hidden Markov Model (HMM) and the Forward channel Performance Index (FCPI) is synthesized by another unit by observing the routers output channels. Then the FCPI indicating its channel characteristics is sent to other router's Learner unit through the outgoing channel link.

### 4.1.2 Communication Domain

If the packet arriving at the router is a conventional router packet, it is directed to the communication domain of the router. This packet carries the routing demand. In the cognitive router, the advice from the cognitive domain is also considered for routing decisions and the router administrator unit routes the messages based on the optimal channel selection algorithms.

## 4.2 Learning

Learning is the principled accumulation of knowledge about the network and it can take place through many means in the network by information sharing [3] [5]. Learning is carried out at different time instances based on the data accumulated over a time representing the current and past network conditions. Implementation of learning in the Cognitive router is carried out through Hidden Markov Model (HMM). The inputs, which are coming as 'Forward Channel Performance Index' representing the characteristics of the forward network channel of all neighboring Routers are fed into Hidden Markov Model of cognition segment of a router as an eight bit pattern. The HMM then predicts the behavioral pattern of each forward channel of all the neighboring Routers. By implementing HMM, the future performance factor of the neighboring Routers are realized in a backward position. Thus it stretches the network awareness to better perimeters.

## 4.3 Planning

Planning requires the knowledge of the present status and the future demand. It takes in to account the demand from the user, their end-to-end goals and the properties of services. Planning is based on 1) user and resource expectations 2) Quality of Service (QoS) Request and performance 3) consumer projection and prediction on communication demand. The planner can organize the resources at the specified time in the near future, thus making sure that all services meet their objectives. It works for satisfying the end-to-end user goals, by efficiently utilizing all available network resources. By prioritizing the service goals and their requirements, a prediction on the resource requirement and configuration is created for future network conditions. For example when there is more demand from a particular region of a network for a service of high bandwidth, the resources such as Bandwidth and Buffer capacity are reserved for it in near future. In addition, neighboring Routers are informed about this situation and are made sure that the demands are entertained within the limits of the network's resources available after reserving for special situation. This leads to better situational awareness, where other neighboring routers try to avoid this link or Router, for its forwarding path until network's availability conditions are improved. In addition, costs can be calculated in advance for each of the services and QoS requirement [2] [5]. Through planning, it is made sure that network remains in the stable condition for the near future, balancing the user requirements and availability of resources in the Network by allocation strategies. Hence, Cognitive Network becomes a forward-looking entity rather than one reacting to problems cropping up from time to time.

## 4.4 Reasoning

Based on the learning and planning outputs, decisions can be made. Decision-making uses the inputs from learning and planning unit and then provides an output in the form of a set of actions that could be followed by the communication segment of the current router. The reasoning involves the application of a selected logic system to draw new inferences and beliefs. Reasoning process can translate declarative knowledge into interpretations of observations and decisions for actions. Decision-making supports the advisor to advise the Router Administrator by providing a set of output that can be used in the network to perform better. The set of output conditions given out from HMM based on the current and past knowledge from different time instances related to each forward router are stored in the form of 8 bit. Based on these values of the conditions of next level of Routers and its forward channels at the next hop-perimeter, the reasoning unit can take decision.

## 5.**Forward Channel Performance Index Synthesizer**

The Channel properties for a router can be characterized by parameters like (i) Bandwidth $P_1$ (ii) delay $P_2$ (iii) jitter $P_3$ (iv) loss $P_4$, that determines the quality of service (QOS). These properties are registered for each channel. Table 1 represents the channel properties for a router $R_n$ under consideration as ($R_n(C_mP_z)$) for channel m (m=1 to 8) and properties/parameters z (z=1 to 4).

For each channel the parameters are computed and if 70% of the channel utilization is realizable at specified demand the evaluator function $f(R_n(C_mP_z))$ is given a value of '1', otherwise it is given a value of '0'.

For the router $R_n$, the calculation is performed for the channel 1 to channel 8 and a bit pattern is formed. For example it can create $f(R_n(C_mP_z))$ = {1 0 1 1 0 1 1 1}, which represents utilization of individual channel of the Router Under Consideration (RUC) like $C_8P_z$,. . . . . . .$C_2P_z$, $C_1P_z$ respectively.

This bit pattern is the Forward Channel Performance Index (FCPI) for that particular router ($R_n$). FCPI always indicates the health of router channels. It summarizes the router performance by index values created using its forward channel Index synthesizer unit as shown in Table 1.

## 6. **Hidden Markov Model (HMM) at the learner unit**

HMM is used to learn the characteristics of the routers placed beyond the current one and predict the condition of the channels there. The advisor of the cognitive segment takes this input and advises the

Table 1. Channel properties at router $R_n$

| Channels | Parameters | | | | E.g. FCPI Result |
|---|---|---|---|---|---|
| | $P_1$ | $P_2$ | $P_3$ | $P_4$ | |
| $C_1$ | $C_1P_1$ | $C_1P_2$ | $C_1P_3$ | $C_1P_4$ | 1 |
| $C_2$ | $C_2P_1$ | $C_2P_2$ | $C_2P_3$ | $C_2P_4$ | 0 |
| $C_3$ | $C_3P_1$ | $C_3P_2$ | $C_3P_3$ | $C_3P_4$ | 0 |
| $C_4$ | $C_4P_1$ | $C_4P_2$ | $C_4P_3$ | $C_4P_4$ | 1 |
| $C_5$ | $C_5P_1$ | $C_5P_2$ | $C_5P_3$ | $C_5P_4$ | 1 |
| $C_6$ | $C_6P_1$ | $C_6P_2$ | $C_6P_3$ | $C_6P_4$ | 0 |
| $C_7$ | $C_7P_1$ | $C_7P_2$ | $C_7P_3$ | $C_7P_4$ | 0 |
| $C_m$ | $C_mP_1$ | $C_mP_2$ | $C_mP_3$ | $C_mP_4$ | 1 |

Router. The router uses this advice and does the routing efficiently and intelligently.

## 6.1 Working of HMM

The Hidden Markov Model (Figure 3.) starts with a finite set of states. Transitions among the states are governed by a set of probabilities (transition probabilities) associated with each state. In a particular state, an outcome or observation can be generated according to a separate probability distribution associated with the state. It is only the outcome, not the state that is visible to an external observer. The states are "hidden" to outside; hence the name Hidden Markov Model [4][9]. The Markov Model used for the hidden layer is a first-order Markov Model, which means that the probability of being in a particular state depends only on the previous state. While in a particular state, the Markov Model is said to "emit an observable" corresponding to that state. One of the goals of using an HMM is to deduce from the set of emitted observables the most likely path in state space that was followed by the system. In our network the state diagram represents the transition occurring at every router. The observables likely path in state space that was followed by the system V1, V2, etc. at every state are the parameters of the channel $P_1,P_{2...}$ etc.

Figure 3. State transition diagram with 3 states.

## 6.2 Hidden Markov Model computation

The model has [4]

- A Set of N states{$S_1,S_2, \ldots S_N$}
- M distinct observation symbols per state $V=\{V_1,V_2,V_3 \ldots V_M\}$.

State transition probability matrix A[ ] where $a_{ij}=P(q_t=j/q_{t-1}=i)$ $1\leq i, j\leq N$ among hidden states.

- Observation probability distribution B(emission of visible states)

$B=\{b_j(k) \mid b_j(k)=P(O_t=v_k \mid q_t=S_j)\ 1\leq k\leq M\ ,\ 1\leq j\leq N\}$ where $O_t$, $q_t$ represent observation and state at time t respectively

- Initial state distribution $P=\{p_i|p_i=P(q_t=i), 1\leq i\leq N\}$

Markov Model can be represented as H={A, B, P}

We consider that some transition occurs from step $t\rightarrow t+1$ (even if it is to the same state) and that some visible symbol be emitted after every step. Thus we have the normalization conditions:

$\sum_j a_{ij}=1$ for all i
$\sum_k b_{jk} = 1$ for all j

where the limits of summations are over all hidden states and all observables, respectively. With these preliminaries, we can now model the Learning Process as follows. Given the coarse structure of a model (the number of states and the number of visible states) but not the probabilities. Given a set of training observations of visible symbols, which arrives at a router at time $t_3$, we can determine the parameters for the next interval $t_4$ by using Viterbi algorithm and system can be trained with modified version of Expectation Maximization (EM) algorithm [4][7][8].

## 6.3 HMM operation

At HMM the cognitive information is collected at selected intervals of time $t_1$, $t_2$, etc.. as shown in Table 2. The timing interval between $t_1$, $t_2$, $t_{p-1}$, $t_p$ is programmable and kept at a low frequency suitable to the network stability.

Time instances $t_1$, $t_{2,...}$.as shown in Table 2. represents the arrival of data into the HMM of a router.

Table 2. Arrival to HMM at Router under consideration (RUC)

| Arrival times of the Cognitive Packets to a Router Under Consideration | | | | |
|---|---|---|---|---|
| | $t_1$ | $t_2$ | $t_{p-1}$ | $t_p$ |
| Router $R_5$ | $R_1$(CmPz) | $R_1$(CmPz) | | |
| | $R_2$(CmPz) | $R_2$(CmPz) | | |
| | $R_3$(CmPz) | $R_3$(CmPz) | | |

Routers $R_1$, $R_2$ etc. shown in Figure 4. are connected to the Router Under Consideration $R_5$.

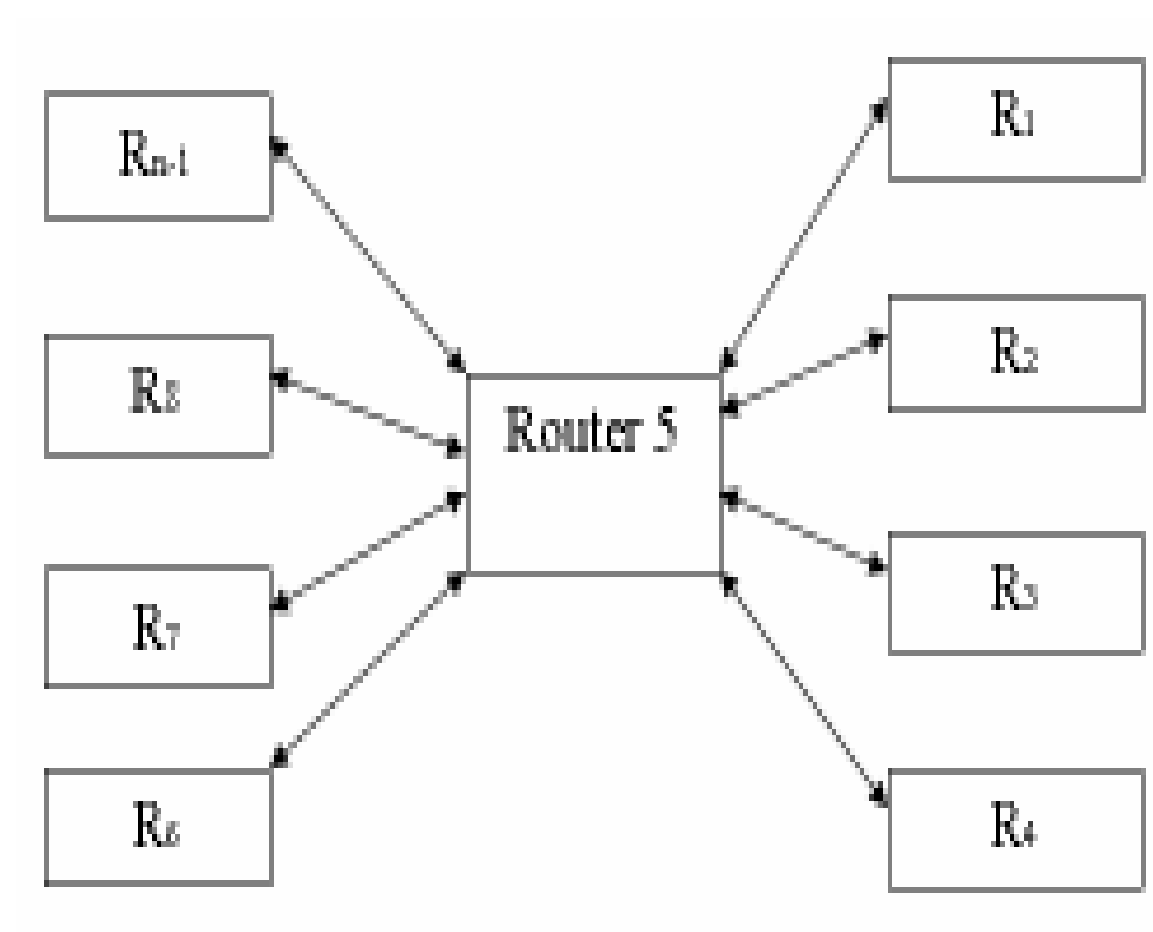

Figure 4. Information to Router $R_5$

At any instant HMM predicts the current index (FCPI) of the forward routers connected to it from the previous state. This information actually represents the status of the outgoing channels of the forward routers connected to it. The HMM collects the FCPI information at intervals $t_1$, $t_2$, etc..from the outside Routers. Table 2. Shows the HMM input values at router $R_5$. For e.g. Routers $R_1$, $R_2$, $R_3$ … are connected to the router $R_5$ through certain channels as shown in Figure 4. Table 3. Shows an example of numerical values of HMM learned output valid at some instance of the intervals.

Table 3. Sample HMM output at Router 5

| Router | Time | | | |
|---|---|---|---|---|
| | $t_1$ | $t_2$ | $t_{p-1}$ | $t_p$ |
| $R_1$ | 153 | 127 | 183 | 254 |
| $R_2$ | 255 | 182 | 127 | 254 |
| $R_3$ | 063 | 153 | 183 | 127 |
| $R_n$ | 225 | 255 | 254 | 252 |

From Table 3. it can be seen that at the instance $t_2$ the HMM computed 153 as FCPI value for forward Router $R_3$. This value is the bit wise representation (153 = 1 0 0 1 1 0 0 1 ), which indicates the status on the outgoing channels of router $R_3$ . Here 153 would mean that the outgoing channels 8, 5, 4 and 1 of router $R_3$ are working good and status on all other channels are 0 means that the lines are not available. Now, a request (router packet) can come to the router $R_5$ from the router $R_4$ to forward the packet to channel 3 of router $R_3$. At router $R_5$, when such a request arrives, the advisor will look into the HMM learned values. Since this value is currently 153, it would mean that all other channels of router $R_3$ except 7, 6, 3 and 2 are utilizable. So the advisor of router $R_5$ would advice its communication domain, of this condition and would give the alternate path. The router then would choose the working channel and does the routing. At an instance $t_p$ the HMM of router $R_5$ will compute a value that gives the state at tp for the router $R_2$. The value at $t_p$ for the router $R_2$ is 254 (1 1 1 1 1 1 1 0) from Table 3. This value means all the outgoing channels of router $R_2$ is functional except channel 1.

The ability of a router to be able to predict the status of the outgoing links of its neighboring routers one level above it, is the feature of 'Stretched Routing Cognition' using HMM. The routing becomes very efficient and intelligent at every router through this approach.

Learning is sufficiently slow process and its result is given to the Advisor along with the output from the planning which takes care of the QoS and policy matters. The Cognitive Domain is incorporated along with the Routing Domain as a part of the architecture of the Router so as to enable successful approach for all Routers in conducting same type of Cognition activities independent of the position of the node and characteristics of the channel available around it. Every Router in the network will be exactly the same as any other in any part of the network and will start learning immediately when this is put into use.

## 7. Conclusion

This paper has described the implementation of an efficient Cognitive Network, which works with learning and prediction capability of its surroundings using HMM. The learning, combined with reasoning works well in the Cognitive Domain along with the routing functions of Routing Domain in a Cognitive Router. It creates a stable network, intelligent enough to look forward to the next hop point layers successfully for reconfiguration of the path offering much better delivery assurance compared to blind routing as prevalent in conventional devices. The planning unit in the cognitive domain advises the router's administrator about the future level of QoS and the service types expected from other routers. The major advantage in this approach is the extremely low level of cognitive packet transport in the network enabling almost full capacity utilization of channels while having a high degree of cognition about the current network state at each node.